\documentclass[10pt]{elsart}
\begin{document}

\begin{frontmatter}

{\rightline{CERN-TH/2001-154}}
{\rightline{hep-ph/xxxxxx}}

\title{\bf{The Symmetry behind Extended Flavour Democracy and Large
Leptonic Mixing}}

\author{G.C. Branco\thanksref{gustavo}}
\address{Theoretical
Physics Division, CERN, 
CH-1211 Geneva 23, Switzerland}

\author{J.I. Silva-Marcos\thanksref{juca}}
\address{Instituto Superior T\'ecnico,
Departamento de F{\'\i}sica,
Centro de F{\'\i}sica das Interac\c c\~oes Fundamentais,
Av. Rovisco Pais, 1049-001 Lisboa, Portugal}

\thanks[gustavo]{Gustavo.Branco@cern.ch; On leave from Instituto Superior T\'ecnico} 
\thanks[juca]{Joaquim.Silva-Marcos@cern.ch}

\begin{abstract}
We show that there is a minimal discrete symmetry which leads to the extended flavour democracy scenario constraining the Dirac
neutrino, the charged lepton and the Majorana neutrino mass term ($M_R$) to
be all proportional to the democratic matrix, with all elements equal. In
particular, this discreet symmetry forbids other large contributions to $M_R$%
, such as a term proportional to the unit matrix, which would normally be
allowed by a $S_{3L}\times S_{3R}$ permutation symmetry. This feature is crucial in
order to obtain large leptonic mixing, without violating 't Hooft's
naturalness principle.
\end{abstract}
\end{frontmatter}

\section{Introduction}

The understanding of the observed pattern of fermion masses and mixings
continues being one of the fundamental open questions in particle physics.\
This flavour puzzle has become even more intriguing with the recent neutrino
data pointing towards neutrino oscillations, with large mixing required in
order to account for the atmospheric neutrino data \cite{atmodata}. In the
absence of a fundamental theory of flavour, one is tempted to consider
specific patterns for the fermion mass matrices which could reflect the
existence of a family symmetry at a higher energy scale \cite{patterns}. The
pattern of fermion masses and mixings may thus provide a valuable insight
into the physics beyond the Standard Model (SM).

One of the most attractive patterns for the quark mass matrices follows from
the suggestion \cite{S3} that there is a $S_{3L}^q\times S_{3R}^u\times
S_{3R}^d$ family permutation symmetry acting on the left-handed quark
doublets, the right-handed up quarks and the right-handed down quarks,
respectively. This family permutation symmetry automatically leads to quark
mass matrices $M_u$, $M_d$ proportional to the so-called democratic mass
matrix \cite{demo}, which has all elements equal to the unity. In the
democratic limit, only the third generation acquires mass and the
Cabibbo-Kobayashi-Maskawa (CKM) matrix is the unit matrix. This is an
interesting result since experimentally one knows that there is a strong
hierarchy in the value of the quark masses, with the first two generations
of quarks much lighter than the third one. Furthermore, the experimentally
observed CKM matrix is close to the unit matrix, as suggested by the
underlying $S_{3L}^q\times S_{3R}^u\times S_{3R}^d$ family symmetry. The
first two generations acquire non-vanishing masses and a non-trivial CKM
matrix is generated when the permutation symmetry is broken.

One may be tempted to extend the above scenario to the leptonic sector and
assume that there is a $S_{3L}^l\times S_{3R}^{cl}$ symmetry acting on the
lepton doublets and the right-handed charged leptons, respectively. If one
pursues this idea, one is confronted with the problem of generating large
leptonic mixing, without violating 't Hooft's naturalness principle \cite
{hooft}. For simplicity, let us assume for the moment the SM without
right-handed neutrinos. Obviously, the $S_{3L}\times S_{3R}$ symmetry leads
to a charged lepton mass matrix proportional to the democratic matrix, which
we denote by $\Delta $. However, as it has been previously pointed out \cite
{yanagi}, the most general effective Majorana mass matrix, allowed by the
permutation symmetry is of the form $a\Delta +b{1\>\!\!\!{\rm I}}$, where
one expects $a$ and $b$ to be of the same order of magnitude. It follows
then that independently of the ratio $a/b$ (provided neither $a$ nor $b$
vanish), both the charged lepton mass matrix and the effective Majorana
neutrino mass matrix are, in leading order, diagonalized by the same unitary
matrix. As a result, in leading order, the leptonic mixing matrix will be
given by the unit matrix. Clearly, no large angles (to solve the atmospheric
neutrino problem, at least) can be generated by a small breaking of the $%
S_{3L}\times S_{3R}$ symmetry.

In the literature, within the framework of democratic mass matrices,
examples with large lepton mixing have been given \cite{fritzsch}, by making
the ad-hoc assumption that the coefficient $a$ vanishes, which is not
dictated by the permutation symmetry. More precisely, the Lagrangean does
not acquire any new symmetry in the limit where $a$ vanishes and therefore
setting $a=0$ clearly violates 't Hooft's naturalness principle. In our
discussion, we have so far restricted ourselves to the case where only
left-handed neutrinos are introduced. We will show in the sequel that
analogous arguments also apply to the case where right-handed neutrinos are
introduced and an effective left-handed Majorana mass matrix is generated
through the seesaw mechanism.

In this paper we shall address the question of whether it is
possible to generate large leptonic mixing using democratic-type mass
matrices, without violating 't Hooft's naturalness principle. We'll show
that this is indeed possible, provided we do not use a $S_{3L}\times S_{3R}$
symmetry, but rather a $Z_3$ symmetry, which is imposed to the quark and
lepton sectors. The $Z_3$ symmetry constrains all fermion mass matrices to
be proportional to the democratic matrix $\Delta $, and a small perturbation
of the symmetry can lead to a correct fermion spectrum and pattern of
mixings. In particular, one may obtain, through the seesaw mechanism, large
mixing in the leptonic sector, without violating 't Hooft's naturalness
principle. The idea of extended flavour democracy (EFD), where the mass matrices of all fermions (i.e., including up and down quarks, charged leptons and neutrinos) are proportional to the democratic matrix has been previously suggested in the literature, as a phenomenological ansatz \cite{egjj}. In this paper, we will show that there is an underlying symmetry which leads to the EFD scenario.

\section{$S_{3L}\times S_{3R}$ symmetry, naturalness and large leptonic
mixing}

For simplicity, let us consider the three generation SM, with the addition
of three right-handed neutrinos. The most general gauge invariant leptonic
Yukawa interaction and mass terms contained in the Lagrangean, can be
written as:

\begin{equation}
\label{lagran}-{L}=Y_l^{ij}\ \overline{L}_i\,\ \phi \,\,\
l_{jR}+Y_D^{ij}\ \overline{L}_i\,\ \tilde\phi \,\,\ \nu _{jR}+\frac 12\,\nu
_{iR}^T\,C\,(M_R)^{ij}\,\nu _{jR}+h.c.\,, 
\end{equation}
where $L_i$, $\phi $ denote the left handed lepton and Higgs doublets, and $%
l_{jR}$, $\nu _{jR}$ the right handed charged lepton and neutrino singlets.
The right-handed Majorana mass term is $SU(3)\times SU(2)_L\times U(1)$
invariant and therefore it is not protected by this symmetry. As a result, $%
M_R$ is naturally large, of the order of the cutoff scale of the low-energy
theory. After spontaneous symmetry breaking, we obtain the mass matrix for
the charged leptons $M_l=<\phi >\ Y_l$ and, besides the Majorana mass $M_R$,
also a Dirac mass matrix for the neutrinos $M_D=<\phi >\ Y_D$.

Imposing a $S_{3_L}\times S_{3_R}$ symmetry on the family structure of this
Lagrangean and choosing, as usual, for the left as well as for the right
handed leptons, real representations of this symmetry, the following
textures for mass matrices are obtained: 
\begin{equation}
\label{demo}
\begin{array}{lll}
M_l=\lambda ^{\prime }\ \Delta \quad ;\quad & M_D=\lambda \ \Delta \quad
;\quad & M_R=\mu \ \left( \Delta +a\ {1\>\!\!\!{\rm I}}\right) 
\end{array}
\end{equation}
where $\Delta $ is the democratic mass matrix with all matrix elements equal
to $1$: 
\begin{equation}
\label{demod}\Delta =\left[ 
\begin{array}{lll}
1 & 1 & 1 \\ 
1 & 1 & 1 \\ 
1 & 1 & 1 
\end{array}
\right] 
\end{equation}
It is important to notice that for the right handed heavy neutrino Majorana
mass matrix, the symmetry does not forbid the existence of the extra term $%
a\ {1\>\!\!\!{\rm I}}$, which, of course, will be of the same order as $%
\Delta $. In the Lagrangean in Eq. (\ref{lagran}), this term is allowed
because Majorana mass terms involve only neutrino fields of the same
chirality. This implies that for the $M_R$ mass term only the $S_{3R}$
symmetry is relevant. When $S_{3_L}\times S_{3_R}$ is broken, the matrices
in Eq. (\ref{demo}) will each acquire an extra small mass term, 
\begin{equation}
\label{demo1}
\begin{array}{ll}
M_l=\lambda ^{\prime }\ \left( \Delta +\varepsilon _l\ P_l\right) \quad
;\quad & M_D=\lambda \ \left( \Delta +\varepsilon _D\ P_D\right) \\  
&  \\ 
M_R=\mu \ \left( \Delta +a\ {1\>\!\!\!{\rm I}}+\varepsilon _R\ P_R\right) &  
\end{array}
\end{equation}
and the effective neutrino mass matrix will be 
\begin{equation}
\label{eff}
\begin{array}{l}
M_{eff}=-M_D\ M_R^{-1}\ M_D^T= \\  
\\ 
=-\frac{\lambda ^2}\mu \ \left( \Delta +\varepsilon _D\ P_D\right) \cdot
\left( \Delta +a\ {1\>\!\!\!{\rm I}}+\varepsilon _R\ P_R\right) ^{-1}\cdot
\left( \Delta +\varepsilon _D\ P_D^T\right) 
\end{array}
\end{equation}
In the sequel, we shall evaluate $M_R^{-1}$ explicitly. One should emphasize
the fundamental difference between the case where $M_R$ is of the form $%
\Delta +\varepsilon _R\ P_R$ and the case where $M_R=\Delta +a\ {1\>\!\!\!%
{\rm I}}+\varepsilon _R\ P_R$. In the first case, $M_R$ is nilpotent and it
does not have an inverse in the limit $\varepsilon _R\rightarrow 0$. As a
result, when $\varepsilon _R$ is small but non-vanishing, the contribution
of the $\varepsilon _RP_R$ term to $M_R^{-1}$ is very large. In the case of $%
\Delta +a\ {1\>\!\!\!{\rm I}}+\varepsilon _R\ P_R$, the situation is quite
different, because, due to the large extra term $a\ {1\>\!\!\!{\rm I}}$, it
has indeed an inverse when $\varepsilon _R\rightarrow 0$ and thus, only a
small term of the same order in $\varepsilon _R$ will appear in $M_R^{-1}$.
Therefore, in the analysis of $M_{eff}$ given in Eq. (\ref{eff}), it is safe
to study the qualitative features of the mass spectrum and neutrino mixing
by putting $\varepsilon _R=0$. As we have argued above, this will not change
qualitatively our results and will allow us to obtain and exact analytical
form for $M_R^{-1}$. Noting that: 
\begin{equation}
\label{inv}\left( \Delta +a\ {1\>\!\!\!{\rm I}}\right) ^{-1}=\frac{-1}{a(3+a)%
}\ \left( \Delta -(3+a)\ {1\>\!\!\!{\rm I}}\right) 
\end{equation}
we find, working out the product in Eq. (\ref{eff}), for the effective
neutrino mass matrix ($\varepsilon _R=0$): 
\begin{equation}
\label{eff1}M_{eff}=\lambda ^{\prime }\ \left( \Delta +\varepsilon _D\
P_D^{\prime }\right) \quad ;\quad P_D^{\prime }=\frac 13\left[ \Delta \
P_D^T+P_D\ \Delta +o(\varepsilon _D)\right] 
\end{equation}
where $\lambda ^{\prime }=-\lambda ^2/\mu (3+a)$. Of course, for $%
\varepsilon _R\neq 0$ a term of the order $\varepsilon _R$ will be added to
this matrix, but it will not change its form or its qualitative features.

It is clear that, in the context of the $S_{3L}\times S_{3R}$ symmetry, the
lepton mixing matrix will either be close to ${1\>\!\!\!{\rm I}}$, or have
only a significant mixing angle in the $(1,2)$ sector. In the $(2,3)$ sector,
the mixing angle will be very small (contrary to what is required by the
atmospheric neutrino data), because both the effective neutrino and the
charged lepton mass matrices have the same texture, namely $\Delta
+\varepsilon \ P$. Thus, in order to have a large leptonic mixing angle in
the $(2,3)$ sector it is crucial that the $a\ {1\>\!\!\!{\rm I}}$ term in
the heavy neutrino Majorana mass matrix is absent. This leads us to the
question: is there a symmetry principle that forbids a large $a\ {1\>\!\!\!%
{\rm I}}$ term in $M_R$, while constraining $M_R$, as well as all other
leptonic mass matrices, to be proportional to $\Delta $? In the next
section, we shall see that such a symmetry does indeed exist.

\section{$Z_3$ symmetry and extended flavour democracy}

Let us consider the Lagrangean of Eq. (\ref{lagran}) and impose a $Z_3$
symmetry realized in the following way: 
\begin{equation}
\label{z3}
\begin{array}{l}
L_i\quad \rightarrow \quad P_{ij}^{\dagger }\ L_j \\ 
l_{iR}\quad \rightarrow \quad P_{ij}\ l_{jR} \\ 
\nu _{iR}\quad \rightarrow \quad P_{ij}\ \nu _{jR} 
\end{array}
\quad ;\quad P=i\omega ^{*}W\quad ;\quad W=\frac 1{\sqrt{3}}\left[ 
\begin{array}{lll}
\omega & 1 & 1 \\ 
1 & \omega & 1 \\ 
1 & 1 & \omega 
\end{array}
\right] 
\end{equation}
where $\omega =e^{i\frac{2\pi }3}$. It can be readily verified that this
indeed a $Z_3$ symmetry since $P^2=P^{\dagger }$, $P^3={1\>\!\!\!{\rm I}}$.
Then, if the Lagrangean is to be invariant, each matrix $M_l$, $M_D$ and $%
M_R $, must obey 
\begin{equation}
\label{sym}P\cdot M\cdot P=M 
\end{equation}
Notice that we do not have $P^{\dagger }\cdot M\cdot P=M$. It is crucial for
our results that Eq. (\ref{sym}) holds and it immediately follows that $\det
(M)=0$, because $\det (P)$ is not real. So $M$ must have, at least, one zero
eigenvalue.

We now prove that $M$ has, in fact, two zero eigenvalues and is always
proportional to the democratic mass matrix $\Delta $. To do this, we write
the unitary matrix $W$ in $P$ as $W=(1/\sqrt{3})\ [\Delta +(\omega -1){%
1\>\!\!\!{\rm I}}]$. It follows then that Eq. (\ref{sym}) is equivalent to 
\begin{equation}
\label{sym1}\omega ^{*}\ \left[ \Delta \ ,\ M\right] =M\cdot \left( \Delta -3%
{1\>\!\!\!{\rm I}}\right) 
\end{equation}
where we have written Eq. (\ref{sym}) in the form $P\cdot M=M\cdot
P^{\dagger }$and used the property $1+\omega +\omega ^{*}=0$. Because $%
\Delta ^2=3\Delta $, if we multiply the right hand side of Eq. (\ref{sym1}),
on the right, by $\Delta $, we get zero and therefore $[\Delta \ ,\ M]\cdot
\Delta =0$, which impliess that:
$$
M\Delta =\frac 13\ \Delta M\Delta 
$$
Subsequently, if we multiply the same equation on the left by $\Delta $ we
find%
$$
\Delta M=\frac 13\ \Delta M\Delta 
$$
Thus $[\Delta \ ,\ M]=0$, but then Eq. (\ref{sym1}) reads 
\begin{equation}
\label{mdelta}M=\frac 13\ M\Delta =\frac 19\ \Delta M\Delta =\lambda \
\Delta 
\end{equation}
where we have used the property $\Delta M\Delta =(\sum M_{ij})\ \Delta $. It
is also easy to check that $P\cdot \Delta \cdot P=\Delta $. Thus \footnote{%
Alternatively, one can derive Eq. (\ref{mdelta}), using Eq. (\ref{sym}),
writing $M=\frac 12\left( P\ M\ P+P^2\ M\ P^2\right) $ and substituting $%
P=(i\omega ^{*}/\sqrt{3})\ [\Delta +(\omega -1){1\>\!\!\!{\rm I}}]$. This
approach can be trivially generalized to a $Z_n$ group.}, 
\begin{equation}
\label{sym2}P\cdot M\cdot P=M\qquad \Leftrightarrow \qquad M=\lambda \
\Delta 
\end{equation}
Therefore, if we impose on the Lagrangean a $Z_3$ symmetry realized in the
way indicated in Eq. (\ref{z3}), all leptonic matrices, $M_l$, $M_D$ and $%
M_R $, are constrained to be of the democratic type, i.e., proportional to $%
\Delta $. In particular, in the limit where the $Z_3$ symmetry holds, $M_R$ will not
contain a term $a\ {1\>\!\!\!{\rm I}}$, since this term is not allowed by
the $Z_3$ symmetry. It can be readily verified that $Z_3$ is the smallest
symmetry which can lead to extended democracy in leptonic mass matrices,
while forbidding the $a\ {1\>\!\!\!{\rm I}}$ term in $M_R$. A $Z_2$ symmetry
would not be sufficient.

\section{Breaking of $Z_3$ and generation of large leptonic mixing}

We shall now investigate what conditions have to be satisfied in order to
achieve large mixing in the leptonic sector, through a small breaking of $%
Z_3 $. Generically, the breaking of $Z_3$ leads to leptonic matrices with
the following form: 
\begin{equation}
\label{pertu}
M_l=\lambda _l\ \left[ \Delta +\varepsilon _l\ P_l\right] \  ;\   
M_D=\lambda \ \left[ \Delta +\varepsilon _D\ P_D\right] \  ;\  
M_R=\mu \ \left[ \Delta +\varepsilon _R\ P_R\right]
\end{equation}
where the $\varepsilon _i\ll 1$ ($i=l,D,R$) and the $P_i$ are of order $1$.
We assume that the perturbation $P_R$ of the right-handed heavy Majorana
neutrinos is such that the inverse of $\Delta +\varepsilon _R\ P_R$ exists.
By noting that one has 
\begin{equation}
\label{dete}\det \left[ \Delta +\varepsilon _R\ P_R\right] =\varepsilon
_R^2\ (x+\varepsilon _R\ y) 
\end{equation}
where $x$ and $y$ are quadratic and cubic polynomials in the (different)
elements $(P_R)_{ij}$, respectively, one readily concludes that $\left[
\Delta +\varepsilon _R\ P_R\right] ^{-1}$is of the form: 
\begin{equation}
\label{inp1}Z\equiv \left[ \Delta +\varepsilon _R\ P_R\right] ^{-1}=\frac 1{%
\varepsilon _R\ (x+\varepsilon _R\ y)}\ \left[ L_0+\varepsilon _R\ X\right] 
\end{equation}
where $L_0$ and $X$ are matrices with respectively linear and quadratic
elements in $(P_R)_{ij}$. Obviously $x,y,L_0$ and $X$ are in general of
order $1$. It is possible to have special cases where either $x$ or $y$
vanish, but not both, since we require that $Z$ exists. Furthermore, it is a
general characteristic of this inverse that $L_0$ and $X$ satisfy the
relations: 
\begin{equation}
\label{lin}
\begin{array}{lll}
\Delta \ L_0=0\quad ;\quad & (\sum X_{ij})=x\quad ;\quad & \Delta \ X\
\Delta =x\ \Delta 
\end{array}
\end{equation}
Applying these algebraic relations to the effective neutrino mass matrix
formula one obtains a transparent formula for $M_{eff}$: 
\begin{eqnarray}
\label{meff}&&M_{eff}=\nonumber\\
&\lambda ^{\prime }& \left[ x\ \Delta
+\frac{\varepsilon
_D^2\ }{\varepsilon _R}\ P_D\ L_0\ P_D
+\varepsilon _D\ \left(\Delta
XP_D+P_DX\Delta +\varepsilon _D\ P_DXP_D\right) \right] 
\end{eqnarray}
where $\lambda ^{\prime }=-\lambda ^2/\mu \left( x+\varepsilon _R\ y\right) $%
.

This expression obtained for the effective
neutrino mass matrix is very important because it tells us when to expect
large mixing for the lepton sector in the case of an aligned hierachical
spectrum for the charged leptons, Dirac and heavy Majorana neutrinos. In
general, i.e., for a generic perturbation $P_R$ in the right-handed Majorana
sector, there will be more than one element $(P_R)_{ij}$ of order $1$, thus implying that the quadratic polinomial $x$ is also of order $1$. So, if the term proportional to $\varepsilon _D^2/\varepsilon _R$
is small, it is clear that the effective neutrino mass matrix will be, just
like the charged lepton mass matrix, to leading order, proportional to $%
\Delta $, and, thus, there will be no large mixing. Therefore, if one wants
to avoid small mixing, one must have that the term proportional to $%
\varepsilon _D^2/\varepsilon _R$, in Eq. (\ref{meff}), be of order $1$ or
larger. Note that, since all $\varepsilon _i$ are of $o(m_2^i/m_3^i)$, large
mixing, and consequently $\varepsilon _R\leq \varepsilon _D^2$, requires
that there is a strong hierarchy (in the sense that the third generation is
much heavier than the first two generations) for the heavy Majorana neutrino
masses. Roughly speaking, if one has that $\varepsilon _D=o(m_c/m_t)$, this
result implies that at least $M_3/M_2=o(m_t^2/m_c^2)$, which is indeed a
very strong hierarchy. The only way to avoid this is to choose perturbations 
$P_R$ in the right-handed Majorana sector, such that $x$ is no longer of
order $1$ but much smaller. The simplest way to do this is by having only
one (diagonal) element of $P_R$ of order $1$. Thus, $x$, which is quadratic
in the different elements of $P_R$, will always be suppressed.

A realization of this scenario was proposed in \cite{egjj} where$\ $all the
perturbation matrices $P_i$ in Eq. (\ref{pertu}) were chosen to be diagonal
matrices such that: 
\begin{equation}
\label{pi}P_i={\rm diag\ }(0,\delta _i,1)\quad ;\quad \delta
_i=o(m_1^i/m_2^i)\quad ;\quad i=l,D,R 
\end{equation}
This choice for the breaking of the family symmetry leads to the following
values for the parameters entering the general expression given by Eq. (\ref
{meff}), 
\begin{equation}
\label{egju}
\begin{array}{llll}
y=0\quad ;\quad & x=o(M_1/M_2)\quad ;\quad & XP_D=0\quad ;\quad & P_D\ L_0\
P_D=P 
\end{array}
\end{equation}
where $P$ is, just like the $P_i$, diagonal with zero first entry. Thus, the
effective neutrino mass matrix 
\begin{equation}
\label{egmeff}M_{eff}=\lambda ^{\prime \prime }\ \left[ \ \Delta +\frac 1x%
\frac{\varepsilon _D^2\ }{\varepsilon _R}\ P\right] 
\end{equation}
with $\lambda ^{\prime \prime }=-\lambda ^2/\mu $, is found to be of the
same form as the Dirac, the right-handed Majorana neutrino and the charged
lepton mass matrices. Furthermore, if the second term in Eq. (\ref{egmeff})
is to be large or of the same order as $\Delta $, the hierarchy of the heavy
Majorana neutrinos will only be $M_3/M_1=o(m_t^2/m_c^2)$, which is a less
pronounced hierarchy\footnote{%
In a somewhat different context, Albright and Barr \cite{albright} used the
seesaw mechanism to generate the large angle solution for the solar neutrino
problem. They were also confronted with a large hierarchy for the
right-handed Majorana neutrinos.}. It was shown that with this specific
perturbation of the extended democracy, one can obtain both an
experimentally acceptable light neutrino spectrum and a pattern of leptonic
mixing in agreement with both the solar and atmospheric neutrino data.

Another example, where a very a strong hierarchy in the right-handed
Majorana neutrino masses is also avoided is the scheme proposed in \cite{jap}%
, where a very special perturbation $P_R$ is assumed, leading to vanishing $%
x $. The effective neutrino mass matrix is then of the form 
\begin{equation}
\label{malbri}M_{eff}=\widehat{\lambda }\ \left[ \frac{\varepsilon _D\ }{%
\varepsilon _R}\ P_D\ L_0\ P_D+\ \Delta XP_D+P_DX\Delta +\varepsilon _D\
P_DXP_D\right] 
\end{equation}
Note however, that such a conspiracy in the sum of quadratic terms in the $%
(P_R)_{ij}$ leading to $x=0$ seems to us likely be unstable with regard to
the renormalization group evolution.

\section{Conclusions}

We have considered a minimal extension of the SM, where the only addition
consists of the introduction of three right-handed neutrinos. We have shown
that if one imposes a $Z_3$ symmetry on the Lagrangean, realized as in Eq. (\ref{z3}), all leptonic mass
matrices, namely the charged lepton mass matrix $M_l$, the Dirac neutrino
mass matrix $M_D$ and the right-handed neutrino Majorana mass matrix $M_R$,
are proportional to the so-called democratic mass matrix $\Delta $. This is
to be contrasted with the situation one encounters when one introduces the
permutation symmetry $S_{3L}\times S_{3R}$. Although this permutation
symmetry leads to $M_l$, $M_D$ proportional to $\Delta $, it allows for a $%
M_R$ containing both a term proportional to $\Delta $ and a term
proportional to the unit matrix. The presence of these two terms in $M_R$
prevents the generation of large leptonic mixing.

On the contrary, in the framework of a $Z_3$ family symmetry, one can obtain
a large leptonic mixing through a small perturbation of the $Z_3$ symmetry.
The fact that large mixing can be obtained through a small perturbation of $%
Z_3$, may seem surprising since, in the democratic limit, there is no
leptonic mixing. The generation of large mixing is due to the fact that one
is perturbing around a singular matrix $\Delta $, where $M_R$ has no
inverse. The possibility of generating large mixing out of small mixing has
already been pointed out in the literature \cite{altarelli}. Obviously, the $%
Z_3$ family symmetry can be trivially extended to the quark sector.

The existence of the $Z_3$ symmetry renders specially appealing the idea of
the EFD scenario, where fermion mass matrices, both in the quark
and lepton sectors, are, to leading order, all proportional to the
democratic matrix. Hopefully, this $Z_3$ symmetry is the low-energy remnant of a larger family symmetry, valid at a higher energy scale.

\section*{Aknowledgments}We would like to thank the Cern Theory Division the kind hospitality extended to us. We are grateful to M.N. Rebelo for a thorough reading of the manuscript. This work received partial support from the Portuguese Ministry of Science - Funda\c c\~ao para a Ci\^encia e Tecnologia, under Project CERN/P/FIS/15184/1999 and Project POCTI/1999/FIS/36288


\begin{thebibliography}{99}


\bibitem{atmodata}SuperKamiokande collaboration, Y. Fukuda et al. Phys. Lett. B 433 (1998) 9; {\it ibid} Phys. Lett. B 436 (1998) 33; {\it ibid} Phys. Rev. Lett. 81 (1998) 1562;; See also M. Vagins, {\it SuperKamiokande's Past, Present and Future}, Recent
Developments in Particle Physics and Cosmology, Proceedings of the NATO ASI
2000, eds. G.C. Branco, Q. Shafi and J.I. Silva-Marcos, Kluwer.

\bibitem{patterns}G. Altarelli, F. Feruglio, Phys. Rep. 320 (1999) 295; Z.
Berezhiani, A. Rossi, hep-ph/0003084; E. Kh. Akhmedov, G. C. Branco, M. N.
Rebelo, Phys. Lett. B 478 (2000) 215; M. Tanimoto, T. Watari, T. Yanagida,
Phys. Lett. B 461 (1999) 345; R. N. Mohapatra, S. Nussinov, Phys. Rev. D 60
(1999) 013002; S. Lola, G. G. Ross, Nucl. Phys. B 553 (1999) 81; J. Ellis,
G. K. Leontaris, S. Lola, D. V. Nanopoulos, Eur. Phys. J. C 9 (1999) 389; C.
Jarlskog, M. Matsuda, S. Skadhauge, M. Tanimoto, Phys. Lett. B 449 (1999)
240; R. Barbieri, G. G. Ross, A. Strumia, JHEP 10 (1999) 020; D. Falcone,
Phys. Lett. B 475 (2000) 92; Q. Shafi, Z. Tavartkiladze, Phys. Lett. B 482
(2000) 145; V. Barger, S. Pakvasa, T. J. Weiler, K. Whisnant, Phys. Lett. B
437 (1998) 107; E. Ma, Phys. Lett. B 442 (1998) 238; B. Stech,
hep-ph/00006076; K. Fukuura, T. Miura, E. Takasugi, M. Yoshimura, Phys. Rev.
D 61 (2000) 073002; T. Miura, E. Takasugi and M. Yoshimura, Phys. Rev.
D 63 (2001) 013001; H. Fritzsch, Z. Xing, Phys. Rev. D 61 (2000) 073016; C.
Wetterich, {Phys. Lett.} {B 451} (1999) 397; J.I. Silva-Marcos, {Phys. Rev.} 
{D 59} (1999) 091301; F. Vissani, hep-ph/9708483; J.A. Casas, V. Di
Clemente, A. Ibarra and M. Quiros, hep-ph/9904295; M. Jezabek and Y. Sumino,
hep-ph/9904382; H.J. Pan and G. Cheng, hep-ph/0102060; N. Haba, J. Sato, M.
Tanimoto and K. Yoshioka, hep-ph/0101334; R. Adhikari, E. Ma and G. Rajasekaran, {Phys. Lett.} {B 486}
(2000) 134; D. Black, A.H. Fariborz, S. Nasri and J. Schechter, {Phys. Rev.} 
{D 62} (2000) 073015; S.M. Barr and I. Dorsner, {Phys. Rev.} {D 61} (2000)
033012; P.Q. Hung, {Phys. Rev.} {D 62} (2000) 053015.

\bibitem{S3}  H. Harari, H. Haut and J. Weyers, Phys. Lett. B 78 (1978) 459.

\bibitem{demo}H. Fritzsch in Proc. of Europhys. Conf. on Flavour Mixing in Weak Interactions (1984), Enice, Italy; G. C. Branco, M. N. Rebelo and J. I. Silva-Marcos, Phys.
Lett. B 237 (1990) 446; G. C. Branco and J. I. Silva-Marcos, {Phys. Lett.} {B  359} (1995) 166; P.M. Fishbane, and  P. Kaus, 
{Phys. Rev.} {D 49} (1994) 3612; {\it ibid}  {Z. Phys.} 
{ C 75} (1997) 1.
359 (1995) 166; G. C. Branco, D. Emmanuel-Costa and J. I. Silva-Marcos,
Phys. Rev. D 56 (1997) 107; P.M. Fisbane and P.Q. Hung, Phys. Rev. D 57 (1998) 2743.

\bibitem{hooft}  G. 't Hooft, {\it Naturalness, Chiral Symmetry and Spontaneous Chiral Symmetry Breaking}, lecture given in Cargese Summer Institute 1979, 135.

\bibitem{yanagi}M. Fukugita, , M. Tanimoto and T. Yanagida, Phys. Rev.
D 57 (1998) 4429

\bibitem{fritzsch}H. Fritzsch, Z.-Z. Xing, Phys. Lett. B 440 (1998) 313.

\bibitem{egjj}  E.Kh. Akhmedov, G.C. Branco, F.R. Joaquim and J.I.
Silva-Marcos, Phys. Lett. B 498 (2001) 237.

\bibitem{albright}  C.H. Albright and S.M. Barr, Phys. Lett. B 461 (1999)
218.

\bibitem{jap}T. Teshima and T. Asai
hep-ph/0009181.

\bibitem{altarelli}G. Altarelli, F. Feruglio, I. Masina, Phys. Lett. B 472 (2000) 382;
\end{thebibliography}
\end{document}